\documentclass[11pt]{article}
\usepackage{graphicx}

\newcommand{\BABARPubYear}    {04}

\newcommand{\BABARProcNumber} {117}
\newcommand{\SLACPubNumber} {10843}
\newcommand{\LANLNumber} {0411047}

\def\babar{\mbox{\slshape {BABAR}}}

\def\Kbar  {\kern 0.2em\overline{\kern -0.2em K}{}\,}

\def\Kz    {\ensuremath{K^0}\,}
\def\Kzb   {\ensuremath{\Kbar^0}\,}
\def\KzKzb {\ensuremath{\Kz \kern -0.16em \Kzb}\,}
\def\Kp    {\ensuremath{K^+}\,}
\def\Km    {\ensuremath{K^-}\,}

\def\KpKm  {\ensuremath{\Kp \kern -0.16em \Km}\,}

\def\Dbar    {\kern 0.2em\overline{\kern -0.2em D}{}\,}

\def\Dz      {\ensuremath{D^0}\,}
\def\Dzb     {\ensuremath{\Dbar^0}\,}
\def\DzDzb   {\ensuremath{\Dz {\kern -0.16em \Dzb}}\,}
\def\Dp      {\ensuremath{D^+}\,}
\def\Dm      {\ensuremath{D^-}\,}

\def\DpDm    {\ensuremath{\Dp {\kern -0.16em \Dm}}\,}

\def\Bbar    {\kern 0.18em\overline{\kern -0.18em B}{}\,}

\def\Bz      {\ensuremath{B^0}\,}
\def\Bzb     {\ensuremath{\Bbar^0}\,}
\def\BzBzb   {\ensuremath{\Bz {\kern -0.16em \Bzb}}\,}
\def\Bu      {\ensuremath{B^+}\,}
\def\Bub     {\ensuremath{B^-}\,}

\def\BpBm    {\ensuremath{\Bu {\kern -0.16em \Bub}}\,}

\def\BorBbar    {\kern 0.18em\optbar{\kern -0.18em B}{}\,}
\def\DorDbar    {\kern 0.18em\optbar{\kern -0.18em D}{}\,}
\def\KorKbar    {\kern 0.18em\optbar{\kern -0.18em K}{}\,}

\mathchardef\Upsilon="7107
\def\Y#1S{\ensuremath{\Upsilon{(#1S)}}\,}

\mathchardef\Deltares="7101
\mathchardef\Xi="7104
\mathchardef\Lambda="7103
\mathchardef\Sigma="7106
\mathchardef\Omega="710A

\def\Deltabar{\kern 0.25em\overline{\kern -0.25em \Deltares}{}\,}
\def\Lbar{\kern 0.2em\overline{\kern -0.2em\Lambda\kern 0.05em}\kern-0.05em{}\,}
\def\Sigbar{\kern 0.2em\overline{\kern -0.2em \Sigma}{}\,}
\def\Xibar{\kern 0.2em\overline{\kern -0.2em \Xi}{}\,}
\def\Obar{\kern 0.2em\overline{\kern -0.2em \Omega}{}\,}
\def\Nbar{\kern 0.2em\overline{\kern -0.2em N}{}\,}
\def\Xb{\kern 0.2em\overline{\kern -0.2em X}{}\,}

\newcommand{\tev}{\ensuremath{\mathrm{\,Te\kern -0.1em V}}\,}
\newcommand{\gev}{\ensuremath{\mathrm{\,Ge\kern -0.1em V}}\,}
\newcommand{\mev}{\ensuremath{\mathrm{\,Me\kern -0.1em V}}\,}
\newcommand{\kev}{\ensuremath{\mathrm{\,ke\kern -0.1em V}}\,}
\newcommand{\ev}{\ensuremath{\mathrm{\,e\kern -0.1em V}}\,}
\newcommand{\gevc}{\ensuremath{{\mathrm{\,Ge\kern -0.1em V\!/}c}}\,}
\newcommand{\mevc}{\ensuremath{{\mathrm{\,Me\kern -0.1em V\!/}c}}\,}
\newcommand{\gevcc}{\ensuremath{{\mathrm{\,Ge\kern -0.1em V\!/}c^2}}\,}
\newcommand{\mevcc}{\ensuremath{{\mathrm{\,Me\kern -0.1em V\!/}c^2}}\,}

\def\mus  {\ensuremath{\rm \,\mus}\,}

\def\mus        {\ensuremath{\,\mu{\rm s}}\,}    

\def\to                 {\ensuremath{\rightarrow}\,}

\def\pep2{PEP-II}

\def\gsim{{~\raise.15em\hbox{$>$}\kern-.85em
          \lower.35em\hbox{$\sim$}~}\,}
\def\lsim{{~\raise.15em\hbox{$<$}\kern-.85em
          \lower.35em\hbox{$\sim$}~}\,}

\,

\def\jetset74   {\mbox{\tt Jetset \hspace{-0.5em}7.\hspace{-0.2em}4}\,}

\long\def\inst#1{\par\nobreak\kern 4pt\nobreak
    {\it #1}\par\vskip 10pt plus 3pt minus 3pt}

\begin{document}
\pagestyle{plain}

\begin{flushright}
SLAC-PUB-\SLACPubNumber \\
\babar-PROC-\BABARPubYear/\BABARProcNumber \\
hep-ex/\LANLNumber \\
November 10,2004 \\
\end{flushright}

\par\vskip 3cm

\begin{center}
\Large \bf Determination of the CKM Matrix Element \boldmath $|V_{cb}|$ 
from Semileptonic \boldmath $B$ Decays
\end{center}
\bigskip

\begin{center}
\large 
Vera G. L\"uth\\
Stanford Linear Accelerator Center \\
Stanford University\\
P.O. Box 20450, Stanford, CA 94309, USA \\
(representing the \babar\ Collaboration)
\end{center}
\bigskip \bigskip

\begin{center}
\large \bf Abstract
\end{center}
We report studies of semileptonic decays, $B \rightarrow X_c \ell\nu$, based on a sample of 88 million $B {\overline B}$ events recorded with the BABAR detector. We have measured four moments of the electron energy distribution and four moments of the hadronic mass distribution, each as a function of the minimum electron energy. From these moments we determine the inclusive branching fraction, the CKM matrix element $|V_{cb}|$, and other heavy quark parameters, using Heavy Quark Expansions (HQE) to order $1/m_b^3$ in the kinetic mass scheme. In addition, we have studied a large sample of exclusive ${B}^0 \to D^{*-}\ell^+\nu$ decays.
This sample is used to extract the vector and axial form factors, the normalization and slope of the HQET form factor to determine $|V_{cb}|$.
\vfill
\begin{center}
Contributed to the Proceedings of the 32$^{th}$ International 
Conference on High Energy Physics, \\
8/16/2004---8/22/2004, Beijing, China
\end{center}

\vspace{1.0cm}
\begin{center}
{\em Stanford Linear Accelerator Center, Stanford University, 
Stanford, CA 94309} \\ \vspace{0.1cm}\hrule\vspace{0.1cm}
Work supported in part by Department of Energy contract DE-AC02-76SF00515.
\end{center}

\section{Introduction}

The CKM matrix element $V_{cb}$, the dominant coupling of the $b$ quark to the charged weak current, is one of the fundamental parameters of the Standard Model.
It determines the rate for $b \rightarrow c \ell \nu$ decays which at the parton level can be calculated accurately. This rate is proportional to $|V_{cb}|^2$ and 
depends also on the quark masses, $m_b$ and $m_c$.

\section{Inclusive Measurements}
To relate measurements of the inclusive semileptonic $B$-meson decay rate 
to $|V_{cb}|$, the parton-level calculations 
have to be corrected for effects of strong interactions.
Heavy-Quark Expansions (HQEs)~\cite{hqe1} have become a useful tool for calculating 
these perturbative and non-perturbative QCD corrections~\cite{hqe2} and for estimating their uncertainties.  
We have chosen the kinetic-mass scheme~\cite{kinetic,kolya} for these expansions in $1/m_b$ and $\alpha_s(m_b)$ (the strong coupling constant). 
To order ${\cal O}(1/m_b^3)$ there are six parameters: the running kinetic masses of the $b$ and 
$c$ quarks, $m_b(\mu)$ and $m_c(\mu)$, and four non-perturbative parameters.
The leading non-perturbative effects
arise at ${\cal O}(1/m_b^2)$ and are parameterized by $\mu_{\pi}^2(\mu)$ and $\mu_{G}^2(\mu)$,  the expectation values of the kinetic and chromomagnetic dimension-five operators. 
At ${\cal O}(1/m_b^3)$, two additional parameters enter, 
$\rho_{D}^3(\mu)$ and $\rho_{LS}^3(\mu)$, the  expectation values 
of the Darwin ($D$) and spin-orbit ($LS$) dimension-six operators. 
These parameters depend on the scale $\mu$ that  separates short-distance from long-distance QCD effects.  

We determine these HQE parameters 
from a fit to the moments of the hadronic-mass and electron-energy distributions in semileptonic $B$ decays to charm particles, $B \rightarrow X_c \ell \nu$, averaged over charged and neutral $B$ mesons. We have measured these moments as functions of $E_{cut}$, a lower limit on the lepton energy (for energy moments we only use electrons, for mass moments we also use muons).  The moments are corrected for detector effects and QED radiation. The charmless contributions are subtracted, using the branching fraction ${\cal B}_{u \ell \nu}$=$(0.22 \pm 0.05)\%~$\cite{btou}.
 
\begin{figure*}[!htb]
\begin{center}
\includegraphics[height=7.5cm]{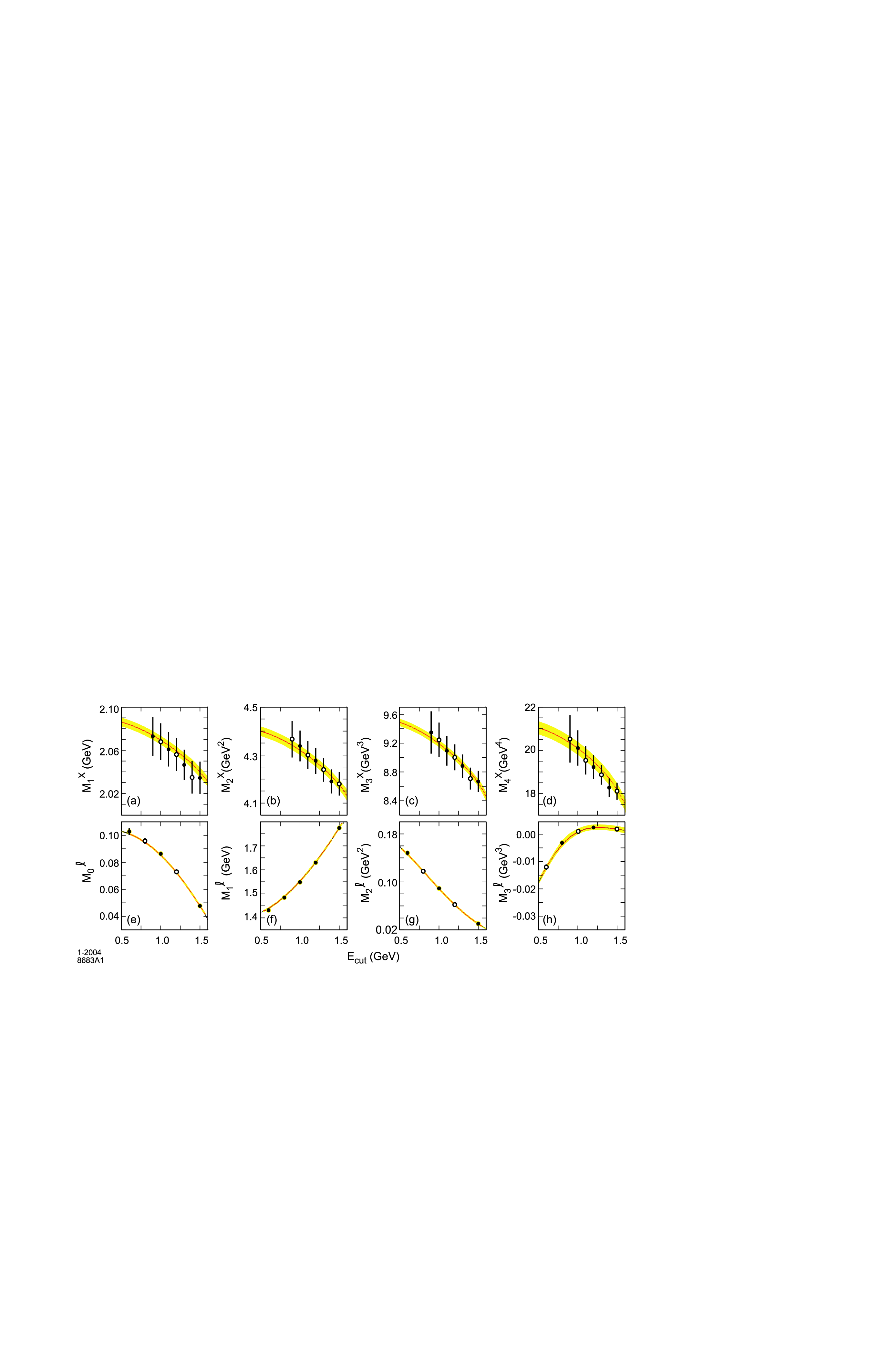}
\caption{
Inclusive $B \rightarrow X_c e \nu$ decays:
The measured hadronic-mass (a-d) and electron-energy (e-h) moments as a function of the cut-off energy, $E_{cut}$, compared with the result of the simultaneous fit (line), with the theoretical uncertainties~\cite{kolya} indicated as shaded bands.  The solid data points mark the measurements included in the fit.  The vertical bars indicate the experimental errors; in some cases they are comparable in size to the data points.  Moments for different $E_{cut}$ are highly correlated~\cite{incl-vcb}.
}
\label{fig:moments}
\end{center}
\end{figure*}

The hadronic-mass distribution is measured in events tagged by the fully reconstructed hadronic decay of the second $B$ meson~\cite{had-mom}.  The hadronic-mass moments are defined as
$M_n^X(E_{cut})=\langle m_X^n \rangle_{E_{\ell}>E_{cut}}$ with $\emph{n}=$ 1,2,3,4.
The electron-energy distribution is measured in events tagged by a high-momentum electron from the second $B$ meson~\cite{e-mom}.  We define 
the first energy moment as $M_1^{\ell}(E_{cut}) = \langle E_{\ell}\rangle_{E_{\ell}>E_{cut}}$  and 
the second and third moments as
$M_n^{\ell}(E_{cut})=\langle(E_{\ell} - M_1^{\ell}(E_{cut}))^n\rangle_{E_{\ell}>E_{cut}}$ 
with $\emph{n}=$ 2,3. 
In addition, we use the partial branching fraction 
$M_0^{\ell}(E_{cut})=\int_{E_{cut}}^{E_{max}}(d{\cal B}_{c \ell \nu}/dE_{\ell})\, dE_{\ell} $.
All measured moments and the results of the least-square fit are shown in Fig.~\ref{fig:moments}. 
The fit describes the data well with $\chi^2=15.0$ for 20 degrees of freedom. For $|V_{cb}|$, the semileptonic branching fraction, and the heavy-quark masses (at the scale of $\mu=1 \gev)$ we obtain~\cite{incl-vcb}
\begin{eqnarray}
\nonumber
|V_{cb}|&=&(41.4 \pm 0.4_{exp} \pm  0.75_{HQE})\,10^{-3}  \nonumber \\
{\cal B}_{c e \nu}&=&( 10.61 \pm 0.16_{exp} \pm 0.06_{HQE})\,\% \nonumber  \\
m_b &=&(4.61 \pm 0.05_{exp} \pm 0.04_{HQE})\,\gev \nonumber \\
m_c &=&(1.18 \pm 0.07_{exp} \pm 0.06_{HQE})\,\gev ,\nonumber
\end{eqnarray}
and $ m_b - m_c=(3.436 \pm 0.025_{exp} \pm 0.018_{HQE} \pm 0.010_{\alpha_s}) \gev$. 
Beyond the statistical, systematic and HQE uncertainties that are included in the fit, 
the limited knowledge of the expression for the decay rate, including various perturbative corrections and higher-order non-perturbative corrections, introduces an additional error on $|V_{cb}|$, assessed to be 1.5\%~\cite{kolya} and included in the stated HQE error.
The choice of the scale $\mu$ is estimated to have a very small impact on $|V_{cb}|$ and the branching fraction~\cite{kolya}.
For the 
non-perturbative parameters in the kinetic-mass scheme up to order $(1/m_b^3)$ we obtain
\begin{eqnarray}
\nonumber
\mu_{\pi}^2 &=& 0.45   \pm 0.04_{exp} \pm 0.04_{HQE}\gev^2 \nonumber \\
\mu_{G}^2   &=& 0.27   \pm 0.06_{exp} \pm 0.04_{HQE}\gev^2 \nonumber \\
\rho_{D}^3  &=& 0.20   \pm 0.02_{exp} \pm 0.02_{HQE}\gev^3 \nonumber \\
\rho_{LS}^3 &=& - 0.09 \pm 0.04_{exp} \pm 0.07_{HQE}\gev^3. \nonumber 
\end{eqnarray}

The fit results are fully compatible with 
independent estimates of $\mu_G^2 = (0.35 \pm 0.07)\gev^2$, based on the $B^* - B$ mass splitting~\cite{kolya}, and of $\rho_{LS}^3 = (-0.15 \pm 0.10)\gev^3$, from heavy-quark sum rules~\cite{sumrules}. The lepton-energy and hadronic-mass moments have a different sensitivity to the fit parameters, but the results for the separate fits are fully compatible with each other and with the global fit to all moments. 
 
Moment measurements and the extraction of $|V_{cb}|$ based on HQEs~\cite{vol-falk} were first performed by the CLEO collaboration~\cite{cleo}. More recently, global fits to a variety of moments were presented~\cite{bauer,cleo2,delphi} using HQEs in different mass schemes. The results presented here are compatible with these earlier fits, but have smaller uncertainties.

\section{Exclusive \boldmath $B^0 \rightarrow D^{*-} \ell^+ \nu$ Decays}

The differential decay rate for the exclusive decay $B^0 \rightarrow D^{*-} \ell^+ \nu$ depends on three helicity amplitudes which are commonly expressed in term of three form factors.  For perfect heavy quark symmetry the three form factor have the same dependence on $w$ and can be related in terms of the Isgur-Wise function. Here $w$ refers to the product of the four-velocities of the $B$ and $D^*$, which corresponds to the relativistic boost of the $D^*$ in the $B$ rest frame, $w=(M^2_B + M^2_{D^*} - q^2)/2 M_B M_{D^*}.$

Exclusive $B^0 \rightarrow D^{*-} \ell^+ \nu$ decays are selected using primarily two kinematic variables, the mass difference $\Delta m= M_{\overline{D}^0 \pi^-} - M_{\overline{D}^0}$
and cos$\theta_{B - D^* \ell}$, where $\theta_{B - D^* \ell}$ represents the angle between the momenta of the $B$ and the $D^* \ell$ pair. 
These distributions are also used to estimate the various background contributions from non-$B {\overline B}$ events, combinatoric and fake backgrounds to the signal $D^*$ and the leptons, and from decays involving higher mass charm mesons states. 
The backgrounds contribute less than 24\% to the selected sample.

\begin{figure*}[!htb]
\begin{center}
\includegraphics[height=3.5cm]{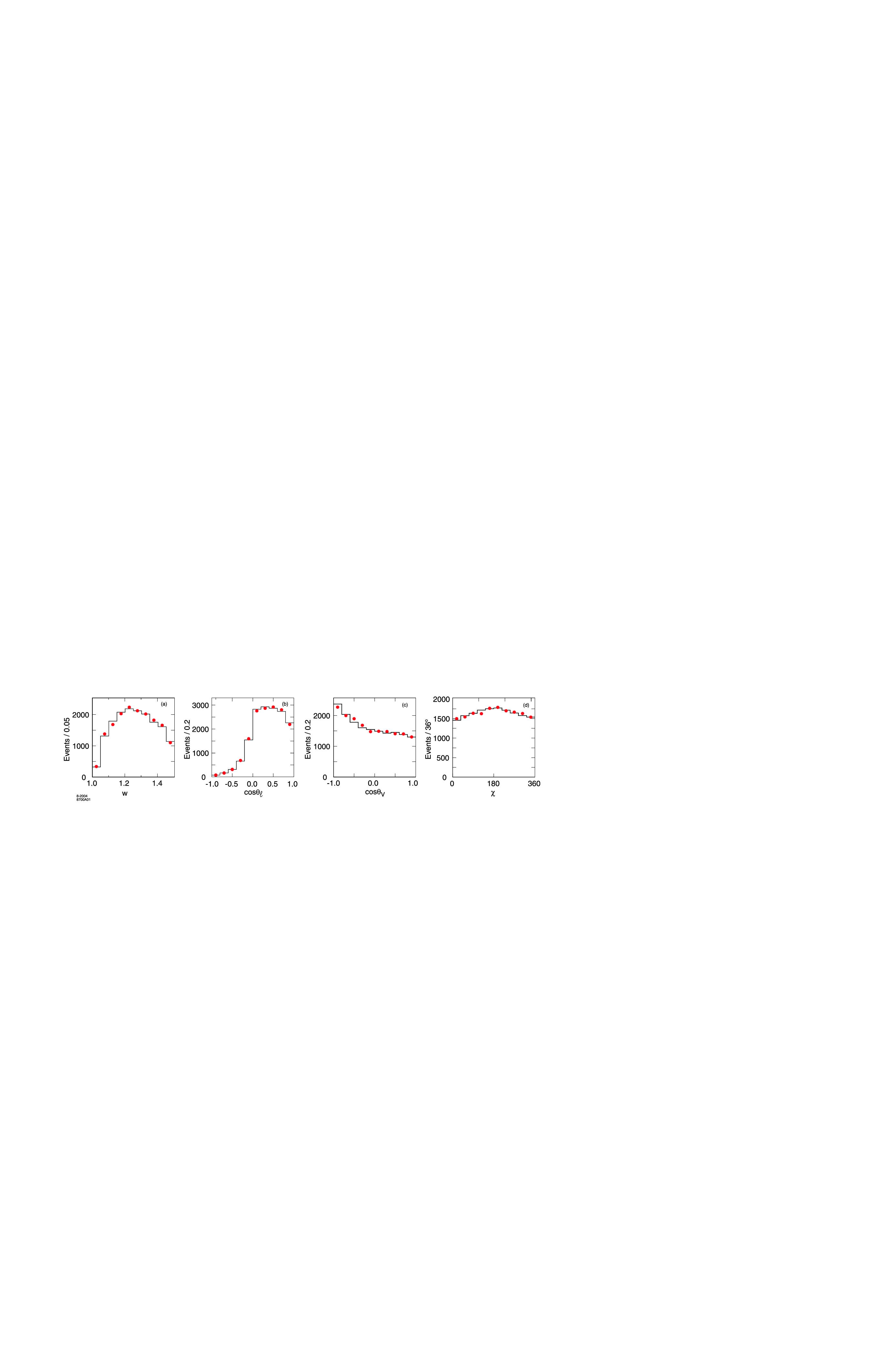}
\caption
{Exclusive $D^{*+} e^- \nu$ decays: Observed one-dimensional decay rates for the four variables a) $w$, b)  
cos $\theta_{\ell}$, and c) cos $\theta_{V}$, the helicity angles for leptons and hadrons, and d)$\chi$, the angle between the decay planes~\cite{babar-ff}.
}
\label{fig:diff-dist}
\end{center}
\end{figure*}

For a sample of about 20,000 selected $B^0 \rightarrow D^{*-} e^+ \nu$ decays (restricted to $\overline{D^0} \rightarrow K^+ \pi^-$ decays) a binned maximum likelihood to the observed four-dimensional decay distribution was performed, see Fig.~\ref{fig:diff-dist}. This fit assumes a linear dependence 
of the form factors on $w$ with a slope $\rho^2$ and form factor ratios, $R_1$ and $R_2$, that are independent of $w$.

The results of the fit~\cite{babar-ff} 
\begin{eqnarray}
\nonumber
R_1(w=1) &=& 1.328   \pm 0.060_{stat} \pm 0.025_{syst} \nonumber \\
R_2(w=1) &=& 0.920   \pm 0.048_{stat} \pm 0.013_{syst} \nonumber \\
\rho^2   &=& 0.759   \pm 0.043_{stat} \pm 0.032_{syst} \nonumber 
\end{eqnarray}
represent a significant improvement over previous measurements~\cite{cleoff}.

The statistical errors still dominate; the largest contribution to the systematic errors is due to the background subtraction.  Though the data do not allow for an independent determination of the $w$-dependence of $R_1$ and $R_2$,
they are compatible with predictions for this dependence by Caprini et al.~\cite{CLN} and others.

To extract $|V_{cb}|$ from exclusive  $B^0 \rightarrow D^{*-} \ell^+ \nu$ decays we integrate over the three decay angles and perform a least-squares fit to the observed $w$ distribution.  For this analysis we use 71,000 decays, including electrons and muons and multiple decay modes of the $D^0$ meson.  We rely on Heavy Quark Effective Theory (HQET) to relate the differential decays rate d$\Gamma/$d$w$ to the product $|V_{cb}| \cdot {\cal A}_1$ at $w=1$. We have adopted the form factor parameterization~\cite{CLN} with
\begin{eqnarray}
\nonumber
R_1(w) &\approx& R_1(1) -0.12(w-1) +0.05(w-1)^2\nonumber \\
R_2(w) &\approx& R_2(1) +0.11(w-1) -0.06(w-1)^2 \nonumber
\end{eqnarray}
and derive the $w$ dependence in terms of a single unknown parameter, $\rho^2_{{\cal A}_1}$,
\begin{eqnarray}
\nonumber
{\cal A}_1(w)/{\cal A}_1(1) 
\approx  1 &-& 8\rho^2_{{\cal A}_1} z +(53\rho^2_{{\cal A}_1}-15)z^2 
           -(231\rho^2_{{\cal A}_1}-91)z^3, \nonumber
\end{eqnarray}
where $z = (\sqrt{w+1}-\sqrt{2})/(\sqrt{w+1}+\sqrt{2})$~\cite{CLN}.

\begin{figure}[!t]
\begin{center}
\includegraphics[height=6cm]{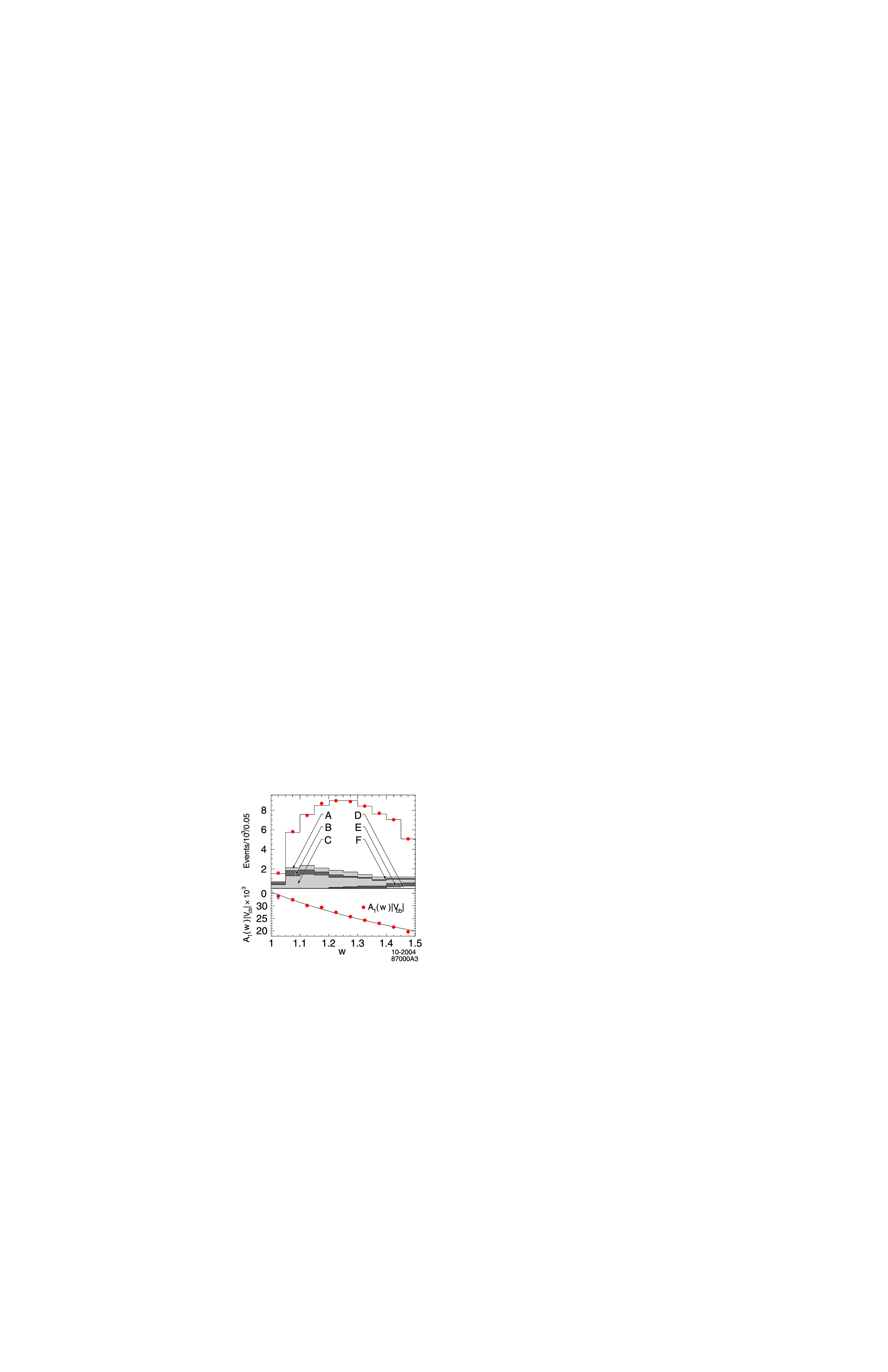}
\caption{
Exclusive $B^0 \rightarrow D^{*-} \ell^+ \nu$ decays: Results of the fit to the decay rate:  Top:  the measured $w$ distribution compared to the fit result~\cite{excl-vcb} (histogram).
Below the signal events the various background contributions are indicated, 
A:~$D^* X \ell \nu$,
B:~uncorrelated $D^* - \ell$ combinations,
C:~fake $D$ decays,
D:~fake leptons,
E:~non-$B{\overline B}$ events ,
F:~correlated $D^* - \ell$ combination.
Bottom: the background- and efficiency-corrected data compared to 
the form factor parameterization with fitted parameters (solid line).
}
\label{fig:vector-rate}
\end{center}
\end{figure}

The data and the result of the fit are shown in Fig.~\ref{fig:vector-rate}.
By extrapolation to $w=1$ we extract~\cite{excl-vcb}
\begin{eqnarray}
\nonumber
{\cal A}_1(1) |V_{cb}| &=& (35.5   \pm 0.3_{stat} \pm 1.6_{syst}) \cdot 10^{-3} \nonumber \\
\rho^2_{{\cal A}_1}  &=& 1.29  \pm 0.03_{stat} \pm 0.27_{syst}. \nonumber 
\end{eqnarray}
Using the lattice calculations for ${\cal A}_1(w)$ at zero-recoil of ${\cal A}_1(w=1)= 0.919 \pm 0.035$ (including a 0.7\% QED correction)~\cite{hashimoto} we obtain
\begin{eqnarray}
\nonumber
|V_{cb}| = ( 38.7 \pm 0.3_{stat} \pm 1.7_{syst} \pm ^{1.5}_{1.3}\, _{{\cal A}1} )
\cdot 10^{-3}. \nonumber
\end{eqnarray}

By integrating over the corrected $w$ distribution we obtain for the branching fraction
\begin{eqnarray}
\nonumber
{\cal B}_{D^{*-} \ell^+ \nu} = ( 4.90\pm 0.07_{stat} \pm 0.36_{syst} ) \%. \nonumber 
\end{eqnarray}
The dominant systematic error is due to the uncertainty in $R_1$ and $R_2$, for which we have used the earlier measurements by the CLEO Collaboration~\cite{cleoff}, 
$R_1(1)=1.18 \pm 0.32$ and $R_2(1)=0.71 \pm 0.21.$ 
These values will be replaced in the future by the more precise BABAR results, taking into account also the $w$ dependence of $R_1$ and $R_2$ including the non-linear terms. 

The measured CKM parameter $|V_{cb}|$ and the exclusive branching fraction  ${\cal B}_{D^{*-} \ell^+ \nu}$ are consistent with earlier 
measurements based on the same parameterization~\cite{vcb}, except for those 
from the CLEO experiment~\cite{CLEOLast}.

\section{Conclusions and Outlook}
With a significantly larger data sample and recent improvements in the theoretical calculation 
the BABAR Collaboration has succeeded in reducing the statistical and systematic error in the determination of $|V_{cb}|$.
The extraction of $|V_{cb}|$ from exclusive decays has still sizable uncertainties.  The current errors are dominated by the form factor uncertainties and the prediction of the decay rate at zero recoil. We expect further improvement, both experimentally and theoretical, in the near future.

\section*{Acknowledgments}
As a member of the \babar\ Collaboration I am indebted to
my \pep2\ colleagues for the excellent operation of the \pep2\ storage rings and to the staff of the associated computing centers for their substantial dedicated effort in support of \babar.
Without their dedication these measurements would not have been possible.

\end{document}